\documentclass[12pt, a4paper]{article}

\pdfoutput=1

\usepackage{amsmath}
\usepackage{array}
\usepackage{amsfonts}
\usepackage{amssymb}
\usepackage{graphicx, rotating}
\usepackage{epsfig}
\usepackage{latexsym}
\usepackage{graphicx}
\usepackage{color}
\usepackage{amsmath,bm,amssymb}
\usepackage{cite}
\usepackage{slashed}
\usepackage{hyperref}
\hypersetup{colorlinks, citecolor=bluscuro, linkcolor=black, urlcolor=bluscuro}
\definecolor{rossos}{cmyk}{0,1,1,0.55}
\definecolor{bluscuro}{rgb}{0.15, 0.2, .85}
\definecolor{bluchiaro}{cmyk}{1,.3,0.,0.1}

\newcommand{\lp}{\left(}
\newcommand{\rp}{\right)}

\newcommand{\be}{\begin{equation}}
\newcommand{\ee}{\end{equation}}
\newcommand{\bea}{\begin{eqnarray}}
\newcommand{\eea}{\end{eqnarray}}

\begin{document}

\begin{titlepage}
\begin{flushright}
DESY 17-227\\
\end{flushright}
\vspace{.3in}

\vspace{1cm}
\begin{center}
{\Large\bf\color{black}
Gravitational radiation  \\ 
\vskip 0.2 cm
from a bulk flow model
}\\
\bigskip\color{black}
\vspace{1cm}{
{\large T.~Konstandin}
\vspace{0.3cm}
} \\[7mm]
{\it {DESY, Notkestr. 85, 22607 Hamburg, Germany}}
\end{center}
\bigskip

\vspace{.4cm}

\begin{abstract}

We perform simulations in a simple model that aims to mimic the hydrodynamic evolution of a 
relativistic fluid during a cosmological first-order phase transitions. The observable we are concerned with is hereby the spectrum of gravitational radiation produced by colliding fluid shells. We present simple parameterizations of our results as functions of the wall velocity, the duration of the phase transition and 
the latent heat. We also improve on previous results in the envelope approximation and compare with hydrodynamic simulations.
\end{abstract}
\bigskip

\end{titlepage}

\section{Introduction \label{sec:intro}} 

It is very  intriguing that gravitational wave observatories like LISA~\cite{eLISA} might be able to probe the sub-atomic world. Nonetheless, several of these links exist: the impact of the QCD equation of state on neutron star mergers~\cite{Annala:2017llu}, gravitational waves from inflation right after the big bang~\cite{Bartolo:2016ami}, gravitational signals from topological defects~\cite{Sanidas:2012ee} or cosmological phase transitions in the early universe~\cite{Caprini:2015zlo}. The last examples are particularly interesting, since -- in principle -- energy scales can be probed that are even beyond the reach of today's particle colliders.

In case a cosmological phase transition is of first order, it proceeds via bubble nucleation. The latent heat is then partially converted into kinetic energy of the Higgs field as well as into a bulk flow of the plasma that fills the early Universe. The contribution to gravitational waves from the Higgs field is probably  well captured by the envelope approximation~\cite{Kosowsky:1992vn} while the contribution from the bulk flow is much harder to quantify. The state-of-the-art calculations are hereby large scale hydrodynamic simulations~\cite{Hindmarsh:2013xza, Hindmarsh:2015qta, Hindmarsh:2017gnf}~\footnote{For a comprehensive review of the field see~\cite{Caprini:2015zlo}.}.

Even though these hydrodynamic simulations are invaluable and shaped our understanding of GW production from phase transitions, they are not without flaws. First, these simulations are very costly and hence typically simulate a relatively small volume with a limited number of bubble nucleations (typically below 100). Often, all bubbles are nucleated at the same time in order to avoid that the first bubble dominates the simulation or even extends beyond the simulation volume.
The high costs also limit the systematic study of the parametric dependence on the wall velocity and the latent heat. 
Another weakness is that the simulation can only cope with relatively small fluid velocities which stems from the general numerical stability of hydrodynamic codes in the non-linear regime. Finally, the dynamic range is constrained by the number of grid points on the lattice. This also prohibits the simulation of highly relativistic bubble wall velocities due to the Lorentz contraction of the Higgs bubble walls. This motivates the search for simplified models as 
the analytic approach in \cite{Caprini:2007xq}, the sound shell model~\cite{Hindmarsh:2016lnk} or the bulk flow model~\cite{Jinno:2017fby}.

The aim of the present work is to present light-weight simulations in a bulk flow model~\cite{Jinno:2017fby} that bypass some of above short-comings. Our analysis is based on the algorithm that was previously used for simulations in the envelope approximation as presented in~\cite{Huber:2008hg}. This approach assumes a vanishing wall thickness and allows for the study of highly relativistic bubble walls. The dynamic range is only limited by numerical accuracy and we present four octaves of frequencies for the gravitational wave spectrum. Ultimately, we want to assess to what extent this simple model can reproduce the results from hydrodynamic simulations.

The plan of the manuscript is as follows. In Section~\ref{sec:method} we describe the model and the general formalism to calculate the GW spectrum. In Section~\ref{sec:algo} we present a more detailed account how the algorithm works that tracks bubble collisions and models the fluid. Our results are presented in condensed form in Sec.~\ref{sec:res} and are discussed in Sec.~\ref{sec:disc}.


\section{Methodology \label{sec:method}} 

In this section, we present the model we employ to mimic the dynamics of the hydrodynamic fluid during (and after) the first-order phase transition. We also review the basic formalism of gravitational wave production and how different histories of nucleating bubbles are generated in our framework.

\subsection{The bulk flow model\label{subsec:model}}

The first simulations of gravitational waves have been performed in the so-called envelope approximation~\cite{Kosowsky:1992rz, Kosowsky:1992vn} (see also~\cite{Kosowsky:1991ua, Huber:2008hg, Weir:2016tov}). The main assumption was hereby that the majority of the energy density resides in the uncollided regions. This was motivated by simulations of the scalar field -- however without the relativistic fluid that filled the Universe at early times.

\begin{figure}[h]
\centering
  \includegraphics[width=0.45\textwidth]{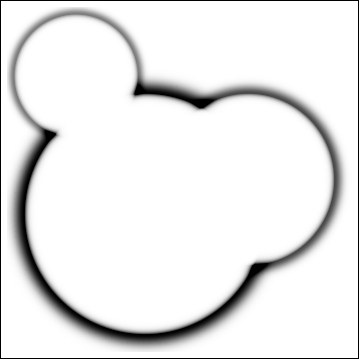}
  \includegraphics[width=0.45\textwidth]{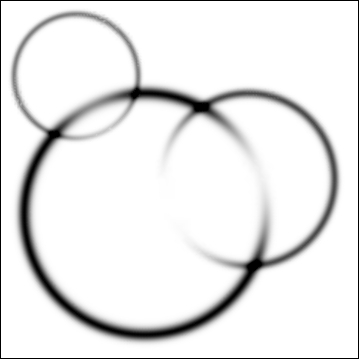}
\caption{\label{fig:sketch}%
\small The plots show two sketches of the envelope (left) and bulk flow (right) approximations. 
}
\end{figure}

In the recent years, it became apparent~\cite{Hindmarsh:2013xza, Hindmarsh:2015qta,  Hindmarsh:2017gnf} that the fluid might actually be the main source of gravitational radiation. The reason being that the sound waves in the fluid persist for a much longer time. Once all initially nucleated bubbles
have collided and percollation is finished, the source in the envelope approximation vanishes, while the fluid still stores most of the latent heat in form of bulk motion.

The main purpose of the present work is to devise a simple model that will mimic this effect and still allows for simplified simulations along the lines of the envelope approximation. As a guiding principle we use hereby energy conservation. Consider a set of bubbles that nucleates at the beginning of the phase transition. 
The bubble walls reach a terminal expansion velocity $v_b$ quickly and we assume  in our model that the bubble walls are infinitely thin.

During the expansion of the uncollided bubble, the latent heat drives the expansion of the bubble. Since the volume of the broken phase scales as $R^3$ ($R$ is the bubble radius) and the surface of the bubble scales with $R^2$, energy conservation requires that the energy density per surface element increases linear with the bubble radius $R$. In contrast, once the bubble surface collided with some surrounding bubbles, no further  latent heat is injected into the fluid. We assume that in this case, the bubble surface continues to propagates with velocity $v_b$ and accordingly the energy density per surface element has to decay as $1/R^2$. In this model, the first collision of a surface element determines at what point the energy density will start to fade away (in the envelope approximation, the anisotropic stress is abruptly removed at this point). Subsequent collisions with other bubble surface elements will not change the dynamics anymore. For the lack of a better name, we call this model in the following the {\em bulk flow model}.

This picture was first put forward in the work~\cite{Jinno:2017fby} and the resulting energy density is sketched in Fig.~\ref{fig:sketch}. The model has clear limitations that have been partially already addressed in~\cite{Jinno:2017fby} and will be part of the discussion below. The main advantage -- compared to the envelope approximation -- is that it captures the fact that the fluid motion sources gravitational waves long after the phase transition is finished. In this respect it is very similar to what is observed in hydrodynamic simulations. The main advantage compared to hydrodynamic simulations is that the simulations are much more economic. This allows to study a wider dynamic range (we will study the frequency range 
$f/\beta \in [10^{-2}, 10^{2}] $) and also simulations with a larger number of bubbles and/or a large number of nucleation histories. We provide results for a wide range of wall velocities $v_b \in [0.01, 1]$.

\subsection{Nucleation histories}

Besides the bulk flow model, we also will improve in some aspects on the results in envelope approximation 
that have been reported in~\cite{Huber:2008hg} and~\cite{Weir:2016tov}.
One main difference between the results here and the ones in~\cite{Huber:2008hg} is that we apply periodic boundary conditions while~\cite{Huber:2008hg} mirrored the bubble configurations at a large spherical boundary bubble. Periodic boundary conditions are also used in the hydrodynamic simulations which makes the results more compatible with this approach. On the other hand, the methods we use below will allow us to only calculate the produced GW spectrum along the three symmetry axes of the system. With a spherical boundary, the GW spectrum along arbitrary directions could be determined. In turn, this means that with periodic boundary conditions, more simulations have to be run or simulations with a larger bubble count. 

\begin{figure}[h]
\centering
  \includegraphics[width=0.65\textwidth]{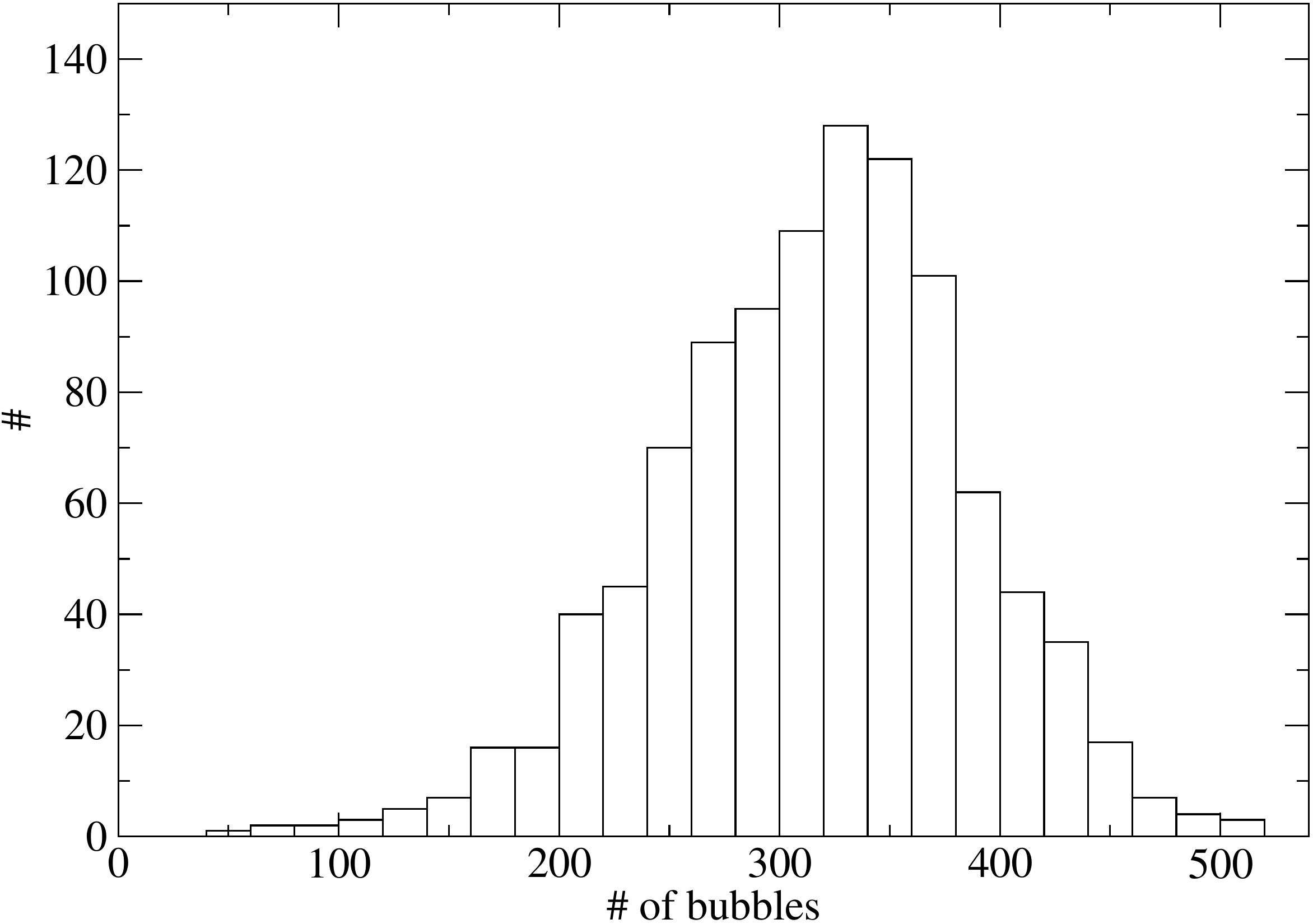}
\caption{\label{fig:histo}%
\small The plot shows the histogram of the distribution  of the bubble count for 1024 nucleation histories in a box of size $(20/\beta)^3$. 
}
\end{figure}

Figure~\ref{fig:histo} shows the distribution of the number of bubbles during the phase transition for $1024$ random  nucleation histories. The probability of nucleating a bubble follows the simple exponential law
\be
\label{eq:def_P}
dP = dt \, \exp(\beta \, t) \, .
\ee
In practice, we choose a time step $dt$ such that $dP < 1\%$ and randomly determine if a bubble is potentially nucleated in this time step or not. The nucleation position is also randomly chosen and the bubble is injected in the simulation in case
the position is still in the symmetric phase.

The volume of the simulation was hereby $V = (20/\beta)^3$ and the average number of bubbles is $316.8$. There is a clear correlation between the number of bubbles nucleated with the time of the first bubble that nucleates which typically happens somewhere in the range $t \in [-2,2]/\beta$. The nucleation of the last bubbles is rather sharply peaked around $t \simeq 7.5/\beta$. Towards the end of the phase transition, the first bubble typically fills $15\%$ of the total volume.

So far, most simulations only chose a handful of  nucleation histories. For example Ref.~\cite{Huber:2008hg} chose eight scenarios while hydrodynamic simulations often only chose a single one for a specific parameter set and/or nucleate all bubbles at the same time. In the present work, we perform a more careful analysis and study a larger number of histories and 
some trends in the prediction of GW spectra. Our strategy to is to randomly chose $3 \times 16$ histories that come in three groups. The first group consists of histories with a number of bubbles in the range $[250,290]$ while the other two groups are drawn from $[290,330]$ and $[330,370]$. We will study trends in the data by comparing the results in the different groups. The final GW spectra are obtained by averaging over the three groups with the weights $0.27$, $0.35$ and $0.38$ which are the weights obtained from the distribution in Fig.~\ref{fig:histo}.

\subsection{Gravitational wave production}

For the determination of the produced GW radiation we follow the 
approach from Refs.~\cite{Kosowsky:1992vn} and~\cite{Huber:2008hg} that is based on Weinberg's master formula~\cite{Weinbergbook}. The energy fraction in GW radiation per octave is given as
\be
\Omega_{GW*} = \omega \frac{dE_{\rm GW}}{d\omega } \frac{1}{E_{\rm tot}} = 
\kappa^2 \left( \frac{H}{\beta} \right)^2 
\left( \frac{\alpha}{\alpha + 1} \right)^2 \, \Delta( \omega / \beta, v_b),
\ee
where we used the definition of the Hubble parameter
\be
H^2 = \frac{8 \pi G \rho_{\rm tot}}{3} = \frac{8 \pi G (\rho_{\rm vac} + \rho_{\rm rad} )}{3} \, ,
\ee
and $\alpha = \rho_{\rm vac}/\rho_{\rm rad}$ denotes the latent heat in units of the radiation energy.

The star indicates that this is the fraction in energy density at the time of production. The energy fraction today is still subject to red-shifting. The dimensionless function $\Delta$ is defined as 
\be
\Delta( \omega / \beta, v_b ) =  \lp \frac{\omega^3}{\beta^3} \rp 
\frac{3 v_b^6 \beta^5}{2 \pi V} 
\int d \hat {\bf k} \, \Lambda_{ij,lm} C^*_{ij} C_{lm} ,
\ee
and the functions $C_{ij}$ contain the geometric information of the  nucleation history and $\Lambda$ projects on the transverse traceless subspace along $\hat {\bf k}$.
It is worth mentioning that for a large volume $V$ (in terms of $\beta^{-3}$) with $N$ bubbles, one obtains
for small wall velocities the scaling
\be
\label{eq:scaling_with_N}
\int d \hat {\bf k} \, \Lambda_{ij,lm} C^*_{ij} C_{lm} 
\propto N \beta^{-8} \propto \frac{V}{v_b^3 \beta^5},
\ee
and hence we recover the well known result~\cite{Kamionkowski:1993fg}
\be
\Omega_{\rm GW*} \propto
\kappa^2 v_b^3 \left( \frac{H}\beta \right) ^2 
\left( \frac{\alpha}{\alpha + 1} \right)^2.
\ee
At this stage it is advantageous to use cylindrical coordinates 
\be
\Lambda_{ij,lm} C^*_{ij} C_{lm} =   ( |C_+|^2 + |C_\times|^2 ) \, , 
\ee
where the two polarizations of the GWs are denoted as $+$ and $\times$ following the standard notation. Finally, the geometry of the nucleation history is encoded in the functions
\bea
\label{eq:def_A_C}
C_p (\omega) &=& \frac{1}{6 \pi}
\sum_n \, e^{ i \omega ( t_n - z_n )}  \int dt \, e^{ i \omega \Delta t_n} 
 \, A_{n,p}(\omega, t), \\
A_{n,p} (\omega,t) &=& \int_{-1}^1 d\zeta 
\, e^{- i v_b \omega \Delta t_n \zeta } B_{n,p}(\zeta,t), 
\eea
and
\bea
\label{def_Bpm}
B_{n,+}(\zeta,t) &=& \frac{(1 - \zeta^2)}2 \int_{S_n} d\phi \, \overline{\Delta t}_n^3 \,  \cos(2\phi), \\
B_{n,\times}(\zeta,t) &=& \frac{(1 - \zeta^2)}2 \int_{S_n} d\phi \, \overline{\Delta t}_n^3 \, \sin(2\phi). 
\eea
Here, we defined the time since nucleation $\Delta t_n = t - t_n$ and the quantity $\overline{\Delta t}_n$ that keeps track of the collided regions of the surfaces of the individual bubbles. For the bulk flow model, we want to weight the surface elements according to 
\be
 \overline{\Delta t}_n = {\rm min}(t-t_n , t_c-t_n) \, ,
\ee
where $t_c(\zeta, \phi)$ denotes the time of the first collision of the surface element. This ensures that the energy density decays as $R^{-2}$ after collision as discussed in Section~\ref{subsec:model}. For the envelope approximation, the energy vanishes after the collision and accordingly, we use
\be
 \overline{\Delta t}_n = (t-t_n) \, \theta_H(t_c-t) \, ,
\ee
using the Heaviside step function $\theta_H$.

Notice that in case of the envelope approximation, the function $\overline{\Delta t}_n$ is piece-wise constant. One only has to keep track of the collided region and can then pull the factor $\overline{\Delta t}_n^3$ out of the integrals $B_{n,p}$ and $A_{n,p}$. In the bulk flow model, the case is more complicated since now the integral over the surface elements are weighted by the time of collision. The feasibility of our algorithm hinges on the fact that the integration over the azimuth angle $\phi$ can be performed analytically. That this is really possible is the topic of the next section.

Furthermore, notice that the energy fraction $\Omega_{GW*}$ contains in principle a double sum over all bubbles. At first sight, it seems strange that this leads to the scaling mentioned in~(\ref{eq:scaling_with_N}).
The reason for this scaling is that the collision region of every bubble is only correlated with the surrounding bubbles. All bubbles that are rather far apart are uncorrelated and their contribution average out in the double sum. This is essential to obtain the scaling~(\ref{eq:scaling_with_N}) and hence a volume independent result for the energy fraction. And this is exactly what the simulation yields.

Finally, let us comment on the velocity dependence in this formalism. The explicit dependence 
on the wall velocity $v_b$ in~(\ref{eq:def_A_C}) is inherited from the physical bubble radius $R_n = v_b \, \Delta t_n$. But there is also an inherent dependence in the nucleation history. Still, it is possible to a large extent to only perform one simulation with bubble wall velocities of the speed of light and then rescale the result to other wall velocities. Since the nucleation history is invariant under a rescaling of space and time, there are two equivalent ways of obtaining the correct scaling. 

First, consider a simulation with bubble wall velocities of the speed of light and then rescale the physical coordinates by a factor $v_b$. In such a simulation, the expansion velocity of the bubble walls is $v_b$. The nucleation history is still valid. The only difference between the two scenarios is that the volume for $v_b < 1$ is smaller which results in an overall factor in the probability (\ref{eq:def_P}) and 
a shift in the time $t_n \to t_n - 3 \log v_b$. This shift is unobservable  
in the final GW spectral density. The rescaling of the spatial coordinates reproduced the correct factor in the exponent of $A_{n,p}$ and also the prefactor $v_b^6 / V \propto v_b^3$. But we also see, that the relative phase between the bubbles should be replaced by
\be
\label{eq:def_rel_phase}
e^{ i \omega ( t_n - z_n )} 
\to
e^{ i \omega ( t_n - v_b \, z_n )}  \, .
\ee
Using this rescaling, one can produce the spectra for different wall velocities easily. In particular, the functions $B_{n,p}$ will not depend on the wall velocities. Moreover, $A_{n,p}$ depends only on the product $v_b \, \omega$. If wall velocities $v_b$ and frequencies $\omega$ are distributed on a logarithmic scale, only a relatively small number of evaluations of $A_{n,p}$ are necessary. In the following we use $60$ values of $v_b \, \omega$ spread over six orders of magnitude (two orders in $v_b$ and four orders in $\omega$). Compare this to $20*40 = 800$ evaluations that would be required without this simplification.

As mentioned before, there is a second way of scaling the simulation, namely scaling in time. Consider a simulation with bubble wall velocities of the speed of light and then rescale the physical time by a factor $t_n \to t_n/v_b$. Again, bubble walls expand with the speed $v_b$ in this rescaled simulation. However, this time the simulation would be a genuinely different simulation since it corresponds to a different bubble nucleation timescale $\beta$. Rescaling the time $t$ and the frequency $\omega$ appropriately, we recover again the formulas (\ref{eq:def_A_C}) and the relative phase (\ref{eq:def_rel_phase}). In particular, the overall factor is $dt^2/\beta^5 \to v_b^3 dt^2/\beta^5$.

In conclusion, we only have to consider a simulation where the bubble walls expand with the speed of light. Spectra for finite wall velocity are obtained using above rescaling.

\section{The algorithm \label{sec:algo}} 

In this section, details of the algorithm are discussed. Readers without interest in the technical details of the numerical implementation can skip to the next section.

\subsection{Geometric considerations}

\begin{figure}[t]
\centering
  \rotatebox{90}{\quad \quad \quad \quad \quad$\zeta = \cos \theta$}
  \includegraphics[width=0.7\textwidth]{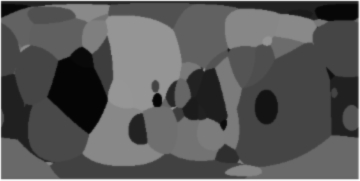}
\hbox{  \hskip 0.2 \textwidth $\phi$ \hskip 0.2 \textwidth } 
\caption{\label{fig:bubble_regions}%
\small The plot shows an example for the data structure that stores which is the first bubble a specific surface element collides with. Different colors correspond to different neighboring bubbles.
}
\end{figure}

We want to evaluate the integrals in (\ref{def_Bpm}) in an efficient way. The main problem is that the collision time $t_c(\zeta,\phi)$ depends on the azimuth angle $\phi$ and the polar angle $\zeta = \cos \theta $ in a nontrivial way. In a first step, we devise a data structure that stores which is the first bubble a specific surface element collides with. An example is shows in Fig.~\ref{fig:bubble_regions}. The data structure is divided into rows with a specific range of the polar angle $\cos \theta$. Every row stores  several (at least one) ordered cells with the information which bubbles are relevant for collision. 
\begin{figure}[h]
\centering
  \includegraphics[width=0.45\textwidth]{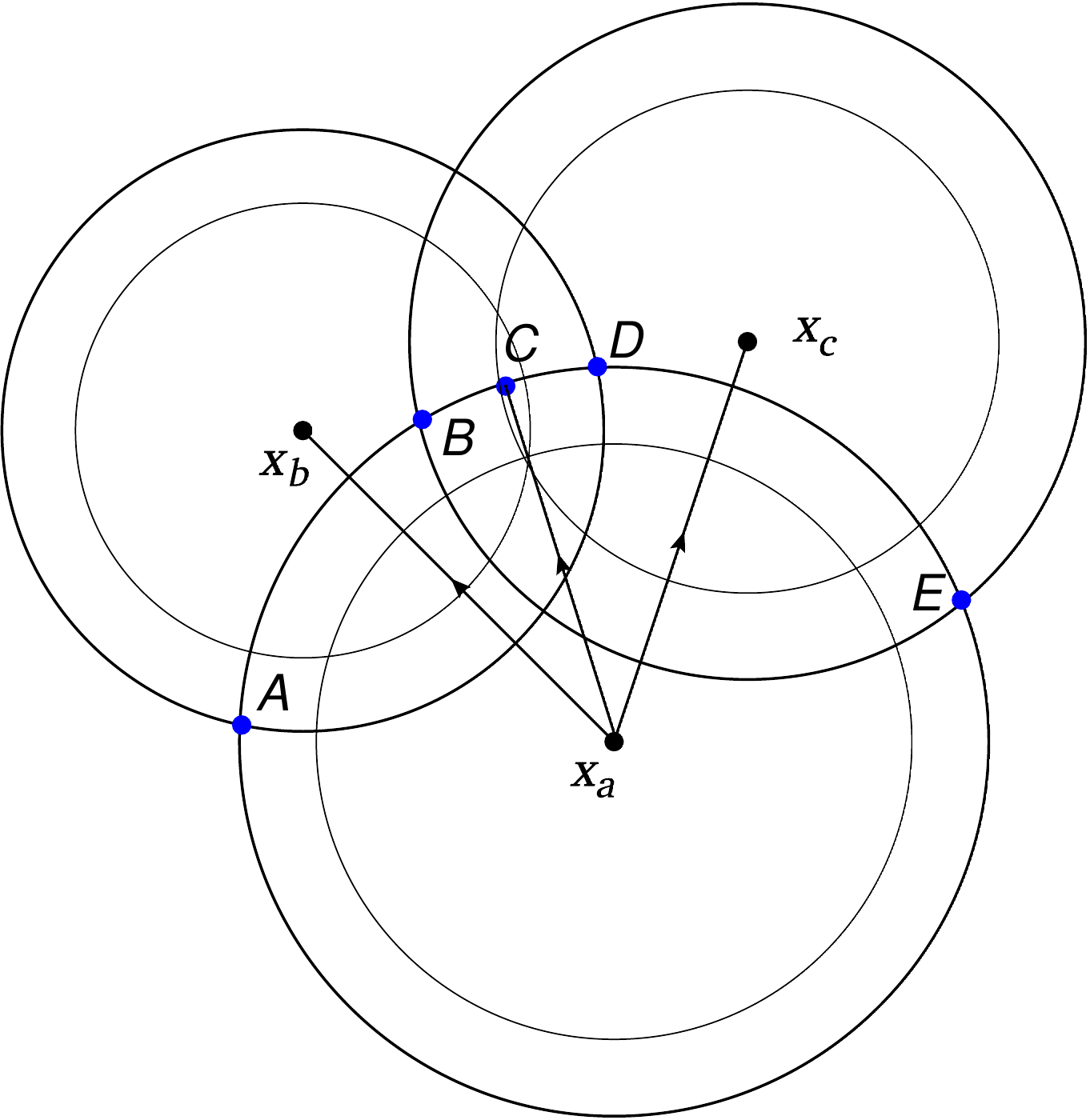}
\caption{\label{fig:geom}%
\small A sketch of the geometry of the problem. The bubble under consideration is positioned at $x_a$ and surrounded by two bubbles at $x_b$ and $x_c$. The four points $A$, $B$, $D$ and $E$ depend on the time $\Delta t_a$ and determine which part of the shell is in the surrounding bubbles according to (\ref{eq:def_constraint}). The two points $A$ and $D$ rely on $\delta^b_\mu = x^\mu_b - x^\mu_a$ while $B$ and $E$ are found using $\delta^c_\mu = x^\mu_c - x^\mu_a$. The point $C$ separates the surface of the bubble at $x_a$ into two regions according to (\ref{eq:earlier_constr}) depending on which bubble the surface element collided first with. For example, points in the segment between $C$ and $B$ are inside both surrounding bubbles but collided with the bubble at $x_b$ before they collided with the bubble at $x_c$. 
}
\end{figure}

In order to generate this data structure, we initialize the data structure with the two mirror images at $z = z_n \pm L$. So the total surface is split into two rows with $\zeta \in [-1,0]$ and $\zeta \in [0,1]$ and the rows contain only the corresponding neighbor in one cell. We then successively add more bubbles to the data structure.

Now, consider two bubbles.  The first one is nucleated at 
$x^\mu_a$ and is the bubble we are currently evaluating in $A_{a,p}$. The second one at $x^\mu_b$ is one of the surrounding bubbles. Consider a volume element $x^\mu$ that is on the surface of both bubbles. Being on the surface of the first bubble implies $|x^\mu-x^\mu_a|^2 = 0$ and hence  
\be
x^\mu = x_a^\mu + \Delta t_a \, (\hat x, 1) 
\equiv x_a^\mu + \Delta t_a \,  \hat X^\mu \equiv x_a^\mu + X^\mu
\ee
where $\hat x$ is an arbitrary normalized direction. Being on the surface of the second bubble then implies
\be
\label{eq:def_constraint}
2 X^\mu \delta^b_\mu = \delta^b_\mu \delta_b^\mu \, , \quad  \delta_b^\mu = x^\mu_b - x^\mu_a \, .
\ee
Hence for a specific direction 
\be
\hat X^\mu = (\hat x, 1)  = (\sin \theta \cos \phi, \sin \theta \sin \phi, \cos \theta,1) \, ,
\ee
on bubble $a$ the time of collision with bubble $b$ can be calculated as
\be
\label{eq:Delt}
\Delta t_a = \frac{\delta^b_\mu \delta_b^\mu}{2 \hat X^\mu \delta^b_\mu} \, .
\ee
Notice that this expression can lead to negative $\Delta t_a$ which is clearly not the solution we are after.
The surface regions of positive and negative $\Delta t_a$ are split by the line $\hat X^\mu \delta^b_\mu = 0$, so $\Delta t_a$ has a discontinuity from $+\infty$ to $-\infty$ there. Luckily, this will never be an issue. The square $\delta_b^\mu \delta^b_\mu$ has to be positive, otherwise the bubble $b$ would be nucleated inside the bubble $a$. Hence, the region that can be potentially covered by bubble $b$ on the surface of bubble $a$ is given by $X^\mu \delta^b_\mu > 0$. 

However, the region that is covered is not determined by this equation but by the requirement that the current bubble collides with the bubble $a$ before the bubbles that are already in the data structure, let's say $c$. One then obtains the constraint
\be
\label{eq:earlier_constr}
\frac{\delta^b_\mu \delta_b^\mu}{2 \hat X^\mu \delta^b_\mu}
< \frac{\delta^c_\mu \delta_c^\mu}{2 \hat X^\mu \delta^c_\mu} \, ,
\ee
that can be rewritten as 
\be
\hat X^\mu c_\mu^{b,c} > 0 \, , \quad 
c_\mu^{b,c} = \delta_\mu^b \, ( \delta^c_\mu \delta_c^\mu ) 
- \delta_\mu^c \, (\delta^b_\mu \delta_b^\mu ) \, . 
\ee
In the following we normalize all constraints $c^\mu$ such that the spatial components are normalized, $\vec c \cdot \vec c = 1$. Hence we can parameterize the constraint as 
\be
c_\mu  = (\sin \bar\theta \cos \bar\phi, \sin \bar\theta \sin \bar\phi, \cos \bar\theta,c_t) \, ,
\ee
that allows for a straightforward evaluation of $\hat X^\mu c_\mu > 0$. For example, the extremal values in terms of the polar angle $\cos \theta$ are given by 
\be
\cos \theta  = \cos \bar \theta \, c_t \pm \sin \bar \theta \sqrt{1- c_t^2} \, .
\ee
Notice that only for $|c_t| \leq 1$ the constraint is non-trivial, since for $c_t < -1$ the constraint is always fulfilled while for $c_t > 1$ never.

Every row of our data structure covers a specific range in terms of the polar angle $\cos \theta$ and the row contains a set of cells. Every cell is bounded by neighboring cells. So if a row contains the set ${b,c,d}$, the boundary of the cells are given by the constraints $c^{b,c}$, $c^{c,d}$ and $c^{d, b}$. Adding a new bubble to the data structure then amounts to checking which cells (and rows) have to be split and which cells are completely covered and can be removed.

\begin{figure}[t]
\centering
  \rotatebox{90}{\quad \quad $\zeta = \cos \theta$}
  \includegraphics[width=0.45\textwidth]{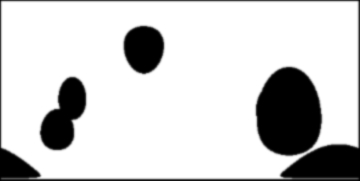}
  \includegraphics[width=0.45\textwidth]{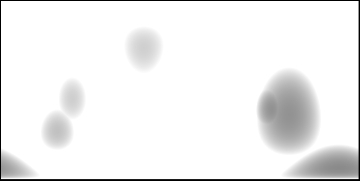}
\vskip 0.05cm
  \rotatebox{90}{\quad \quad $\zeta = \cos \theta$}
  \includegraphics[width=0.45\textwidth]{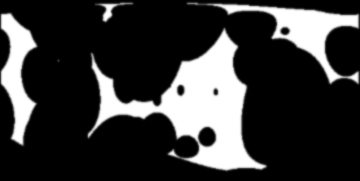}
  \includegraphics[width=0.45\textwidth]{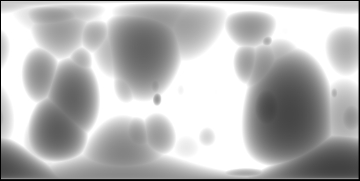}
\vskip 0.05cm
  \rotatebox{90}{\quad \quad $\zeta = \cos \theta$}
  \includegraphics[width=0.45\textwidth]{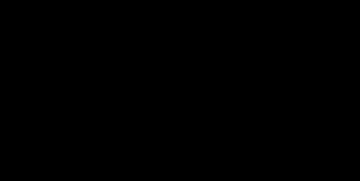}
  \includegraphics[width=0.45\textwidth]{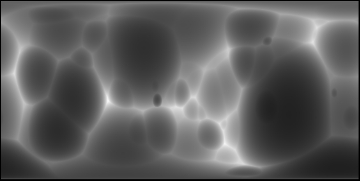}
	\hbox{ $\phi$ \hskip 0.5 \textwidth $\phi$ } 
\caption{\label{fig:densities}%
\small The plot shows the energy density on the surface of the biggest bubble in a simulation for three different times ($5/\beta, 6.4/\beta, 8.2/\beta$). The energy density increases from dark to light regions. The energy density is normalized to the maximal energy density possible at the specific time, {\em i.e.}~black regions denote no energy density and white regions denote uncollided regions. The plots show the envelope (left) and bulk flow (right) approximations. 
}
\end{figure}

This data structure has to be constructed once for all times. This is beneficial, since it will contain only a relatively small number of surrounding bubbles which saves the time to check overlap with all bubbles for all time steps. For a fixed time, the cells in the data structure are eventually surrounded by uncollided regions. For every cell, one has to check, according to (\ref{eq:Delt}) if 
${2 \hat X^\mu \delta^b_\mu} > (\delta^b_\mu \delta_b^\mu)/\Delta t_a$ that can again be transformed into a normalized constraint (using the fact that the time component of $\hat X$ is 1). Using this constraint, rows and cells have to be split again to account for uncollided regions. An example for the resulting energy density in spherical coordinates is shows in Fig.~\ref{fig:densities}

\subsection{Integration over the azimuth angle $\phi$}

The heart of the algorithm is the based on the fact that the innermost integration can actually be performed analytically. In the envelope case this is trivial using 
\bea
\int^{\phi_0}_0 \, d\phi \sin(2\phi) &=& \sin(\phi_0)^2 \, ,\\
\int^{\phi_0}_0 \, d\phi \cos(2\phi) &=& \sin(\phi_0) \cos(\phi_0) \, .
\eea
The case for the bulk fluid model is a bit more involved since an additional factor $\Delta t_n^3$ is contained in the integrand of (\ref{def_Bpm}). According to (\ref{eq:Delt}) all relevant integrals can be brought to the form
\be
I_s(\alpha,\phi_0) \equiv 
\int^{\phi_0}_0 \, d\phi \, \frac{\sin(2\phi)}{[1 + \alpha \cos(\phi)]^3} = 
\frac{2  \, [ 1 + \cos(\phi_0) + 2 \alpha \cos(\phi_0) ] \sin(\phi_0/2)^2}
{ (1 + \alpha)^2 (1 + \alpha \cos(\phi_0))^2 } \, ,
\ee
and
\bea
\label{eq:Ic_def}
I_c(\alpha,\phi_0) &\equiv& 
\int^{\phi_0}_0 \, d\phi \, \frac{\cos(2\phi)}{[1 + \alpha \cos(\phi)]^3}  \nonumber \\
&=&
-\frac{3 \alpha^2}{(1 - \alpha^2)^{5/2}}
{\rm arctan} \left[
\sqrt{\frac{\alpha - 1}{\alpha + 1}} \tan(\phi_0/2) \right] \nonumber \\
&& - \frac{(\alpha (2 + \alpha^2) + (-2 + 5 \alpha^2) \cos(\phi_0) \sin(\phi_0)}
{2 (1 - \alpha^2)^2 (1 + \alpha \cos(\phi_0))^2} \, .
\eea
For $|\alpha| < 1$ the expression $I_s$ is generally valid while the expression for $I_c$ is valid in the domain $\phi_0 \in (-\pi, \pi)$. Even though the integrand is periodic, $I_c$ is only periodic up to a linear term.

For $|\alpha| \geq 1$ the integrands have singularities and the expressions are only valid up to the singularities in the integrand, $\cos(\phi_0) = -1/\alpha$. Using the relation 
\bea
I_c(\alpha,\phi_1) - I_c(\alpha,\phi_0) 
&=&  I_c(-\alpha,\phi_1 + \pi) - I_c(-\alpha,\phi_0 + \pi) \, , \\
I_s(\alpha,\phi_1) - I_s(\alpha,\phi_0) 
&=&  I_s(-\alpha,\phi_1 + \pi) - I_s(-\alpha,\phi_0 + \pi) \, , 
\eea
one can map the desired integrals to the regions that are well behaved. Furthermore, there is an numerical instability in $I_c$ near $\alpha \simeq 1$ due to cancellation between the two terms in (\ref{eq:Ic_def}). As long as the argument of $\rm arctan$ is relatively small, one can expand the expression around $\alpha = 1$ to obtain an approximation that is more stable when evaluated numerically.

\subsection{Remaining integrations}

The two remaining integrals in~(\ref{eq:def_A_C}) are of the form 
\be
\int \, dx \, \exp (i \, \omega \, x) f(x) \, .
\ee
However, we cannot use the standard fast Fourier transform algorithms to evaluate these integrals since $\omega$ is logarithmically distributed in our approach. 

We approximate these integrals with two methods. In the first one, we approximate the function $f(x)$ by a second order polynomial and then perform the integration. This leads to accurate results when $\omega \, dx$ is large. However, when $\omega \, dx$ is small, this leads to numerical instabilities due to cancellations between terms with different inverse powers of $\omega \, dx$. In this case, we approximate the full integrand $\exp (i \, \omega \, x) f(x)$ by a second order polynomial and integrate. Due to the precision we are using ($\simeq 10^{-16}$ for floating point numbers of {\em long double} type) these two methods are used for $\omega \, dx \gtrless 3 \times 10^{-4}$. 

\begin{figure}[b!]
\centering
  \includegraphics[width=0.95\textwidth]{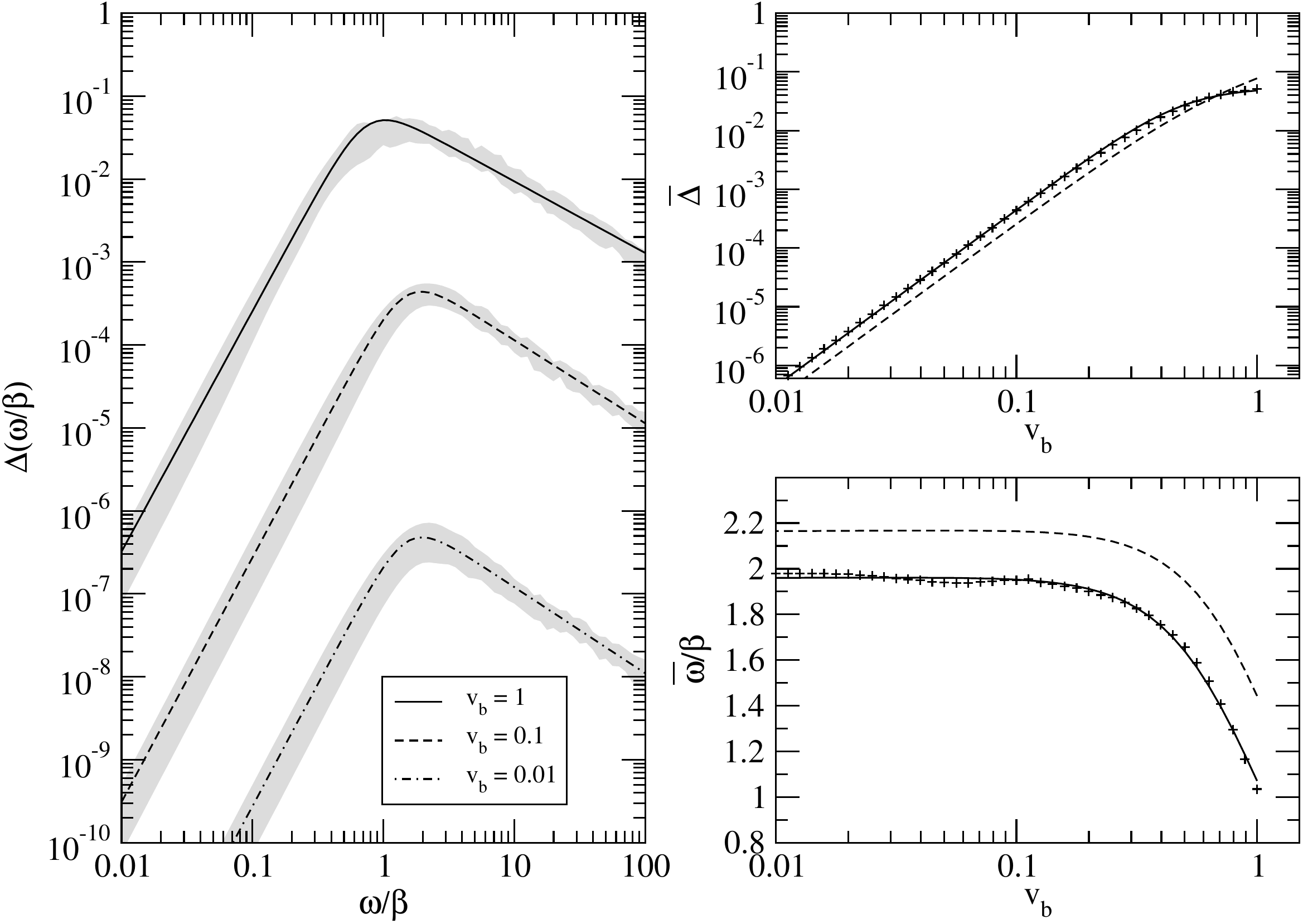}
\caption{\label{fig:sum_env}%
\small GW spectra for the envelope approximation. The left plot shows the cumulative spectra $\Delta(\omega/\beta, v_b)$ for three different velocities, $v_b \in \{ 1, 0.1, 0.01 \}$. The shaded region denotes the variance obtained from averaging over 48 nucleation histories. The right panels show the peak frequency and the peak amplitude as a function of the wall velocity. For comparison, we show the best fit from Ref.~\cite{Huber:2008hg} (dashed) as well as the best fit to the current data (see table \ref{tab:fits} and \ref{tab:fits2}).
}
\end{figure}

We evaluate the integrals for $C_{p}$ and $A_{n,p}$ in an adaptive way. First, we calculate the integrals for $\omega=0$ and further refine the integral until the local relative error is below $10^{-7}$ and $10^{-10}$ respectively. Once this accuracy is reached, the resulting grid of support points are used to evaluate the integral for all values of $\omega$. This high precision is required to obtain meaningful results at the high frequency end of the spectrum.

One peculiarity of the bulk flow model is that the energy density in the surface of the expanding bubbles persists even beyond the end of the phase transition (see the right panels of Fig.~\ref{fig:densities}).
For these times, the function $B_{n,p}(\zeta, t)$ becomes constant in time. The corresponding contribution to  
$C_p$ is proportional to 
\be
\label{eq:infiPiece}
\int dt \, e^{ i \omega \Delta t_n} 
 \, A_{n,p}(\omega, t) = 
\int_{-1}^1 d\zeta \, e^{ i \omega \Delta t_n^{\rm end} (1 - v_b \zeta)} 
\frac{i \, B_{n,p}(\zeta)}{\omega(1 - v_b \, \zeta)} \, .
\ee
In particular, the time integration is finite. This contribution scales as $\omega^{-1}$ for small $\omega$
and as $\omega^{-2}$ for large $\omega$. However, the leading contribution for large $\omega$ in fact cancels against the contribution from the time integration during the phase transition, $\Delta t_n < t< \Delta t_n^{\rm end}$, such that $C_p$ scales as $\omega^{-3}$ for large $\omega$. This serves as another important cross check for the stability of our numerical integration.

\section{Results \label{sec:res}} 

\begin{figure}[t]
\centering
  \includegraphics[width=0.95\textwidth]{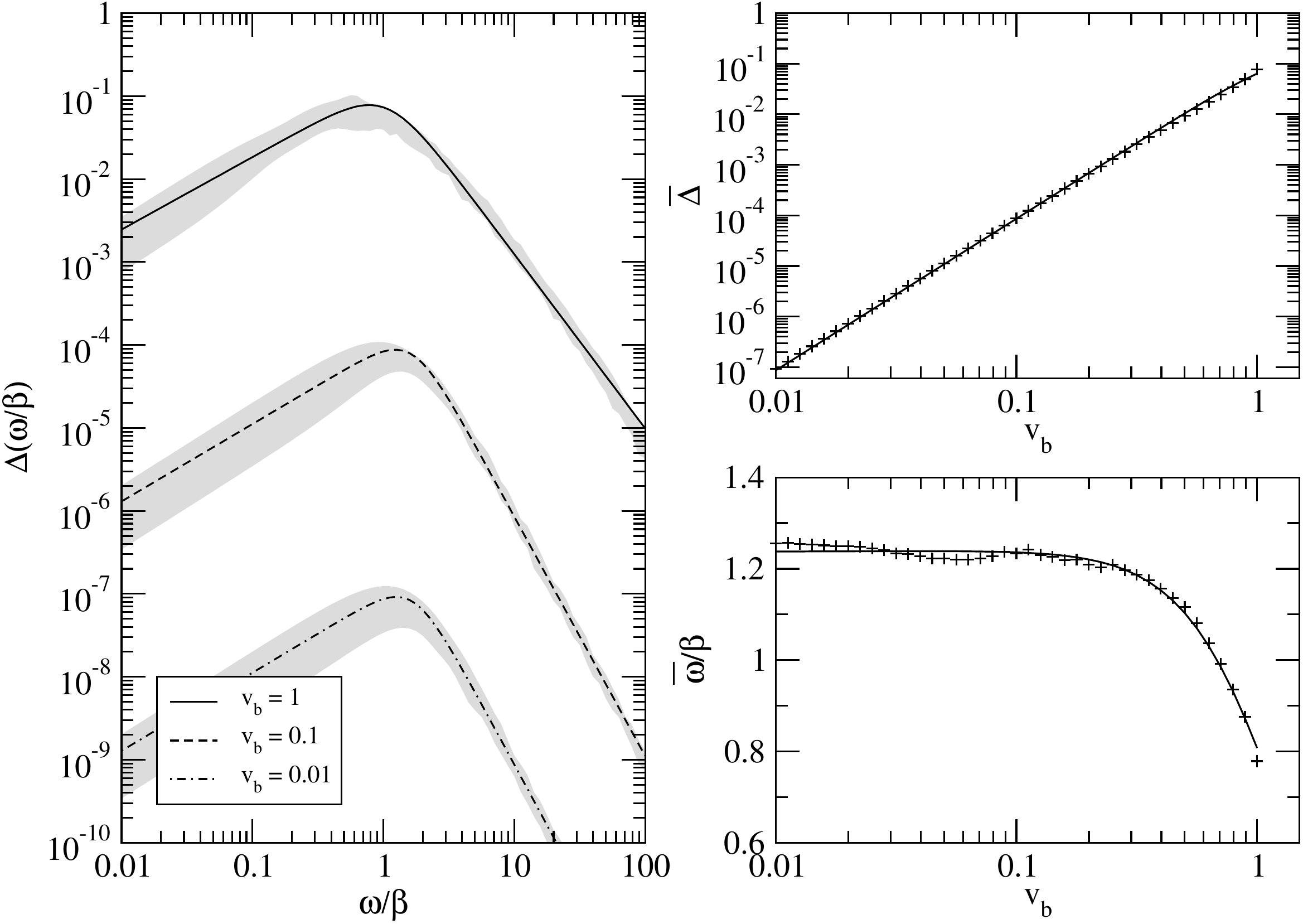}
\caption{\label{fig:sum_fluid}%
\small Same as Fig.~\ref{fig:sum_env} for the bulk flow model. 
}
\end{figure}

In this section, we present the final results of our simulations. As described before, we simulate in total 48  nucleation histories in three batches with different bubble count. The simulations are performed in a box of size $(20/\beta)^3$ which leads to around $\sim 300$ bubbles in each simulation. Using these three batches, one can study if the observed spectrum follows some trends with the bubble count. A strong dependence on the bubble count could distort the result if only  nucleation histories with average bubble count are studied. We did not observe such strong correlations. For a bubble wall velocity of the speed of light, no clear correlation is observed while for small bubble wall velocities, the amplitude seems to be anti-correlated with the bubble count. However, this effect is rather weak (the amplitude is approximately inversely proportional to the bubble count). We weight the three batches according to the distribution in Fig.~\ref{fig:histo} which should take these correlations into account.

The 48 spectra are averaged and we assign an error to the spectrum that results from the variance of this averaging. The results are shown in Figs.~\ref{fig:sum_env} and~\ref{fig:sum_fluid}. 
We parameterize the spectrum as
\be
\label{eq:spec_fit}
\Omega_{\rm GW}(f) = \bar \Omega_{\rm GW} 
\frac{(a+b) \, \bar f^{b} f^a}{ b \bar f^{(a+b)} + a f^{(a+b)}},
\ee
where $\bar f$ and $\bar \Omega_{\rm GW}$ denote the peak frequency and amplitude of the spectrum.
The spectrum observed today is obtained by red-shifting
\bea
\label{res2}
\bar f &=& 2.62 \times 10^{-3} \textrm{mHz} \left( \frac{\bar \omega}{\beta} \right) 
\left( \frac{\beta}{H_*} \right)
\left( \frac{T_*}{100 \textrm{ GeV}}\right) \left( \frac{g_*}{100}\right)^{1/6}, \\
h^2 \bar \Omega_{\rm GW} &=& 1.67 \times 10^{-5} \bar \Delta \,
\kappa^2 \left( \frac{H}{\beta} \right)^2 
\left( \frac{\alpha}{\alpha + 1} \right)^2
\left( \frac{100}{g_*}\right)^{1/3} \, ,
\eea
where $g_*$ denotes the effective number of degrees of freedom in the plasma of the early Universe at the time of production.

The peak amplitude $\bar \Delta$ and frequency $\bar \omega$ also have a dependence on the wall velocity that is shown in the right panels of Figs.~\ref{fig:sum_env} and~\ref{fig:sum_fluid}. The figures also show fits to the data that are given in Tables~\ref{tab:fits} and \ref{tab:fits2}.

\begin{table}
\centering
\setlength\extrarowheight{5pt}
  \begin{tabular}{ | c |  c | c | }
    \hline 
    & $\{ a,b\}$ for $v_b \simeq 1$
    & $\{ a,b\}$ for $v_b \ll 1$ \\ [5pt]
    \hline
    {\rm envelope} & $\{ 2.9, 0.9 \}$ & $\{ 2.95, 1 \}$  \\   [5pt]
    \hline
    {\rm bulk flow}  & $\{ 0.9, 2.1 \}$ & $\{ 0.95, 2.9 \}$  \\ [5pt]
    \hline
  \end{tabular}
\caption{%
\small Parameters for the spectral shape in (\ref{eq:spec_fit}). 
} \label{tab:fits}
\end{table}
\begin{table}
\centering
\setlength\extrarowheight{10pt}
  \begin{tabular}{ | c |  c | c | }
    \hline
    & $\bar \Delta$ & $\bar \omega$ \\  [10pt]
    \hline
    {\rm envelope} 
& $\displaystyle\frac{0.44  \,v_b^3}{1 + 8.28  \, v_b^3}$ 
&  $\displaystyle\frac{1.96}{1 - 0.051  \, v_b + 0.88  \, v_b^2}$  \\ [10pt]
    \hline
    {\rm bulk flow}  
& $\displaystyle\frac{0.0866  \,v_b^3}{1 + 0.354  \, v_b^3}$ 
&  $\displaystyle\frac{1.24}{1 - 0.047  \, v_b + 0.58  \, v_b^2}$  \\ [10pt]
    \hline
  \end{tabular}
\caption{%
\small Fits for the peak quantities in Figs.~\ref{fig:sum_env} and~\ref{fig:sum_fluid}. 
} \label{tab:fits2}
\end{table}

\section{Discussion \label{sec:disc}} 

We presented novel results for the gravitational wave spectrum produced by a simplified model of bulk flow (see Sec.~\ref{subsec:model} for a description of the model). 
Besides, we also provide improved GW spectra for the envelope approximation that also serve as a validation for our code. 

Compared to former analysis~\cite{Huber:2008hg}, the present simulations are performed with periodic boundary conditions. This has the advantage that they are directly compatible with the results from hydrodynamic simulations on a grid. However, with the present method, it is only possible to determine the GW radiation along the three symmetry axes of the box. This makes it harder to disentangle the variation in the GW signal due to the quadrupole nature of the radiation from the variation due to the nucleation history. In particular, the variance in our results most probably overestimates the true error.
 
The present simulation uses 48 scenarios with $\sim 300$ bubbles averaged over six directions. In comparison, the simulation from Ref.~\cite{Huber:2008hg} used eight scenarios with $\sim 100$ bubbles averaged over 32 directions. In particular, the eight simulations have been chosen with a bubble count around $100$ in mind which introduces a certain bias and a systematic error that was not quantified. Overall, the current simulations should be more precise and also give a more realistic estimate of the error.
Numerically, our findings agree well with the former results in the envelope approximation (see Fig.~\ref{fig:sum_env}). We observe a somewhat smaller GW signal for a wall velocity close to the speed of light and somewhat larger signals for small wall velocities. These effects are of order $50\%$ which is well within the uncertainty of our method. The peak frequency is slightly ($\sim 25\%$) smaller than previously reported. 

Concerning the bulk flow model, our results also agree well with the soft tail of the spectrum ($\omega \ll \beta$) that was already presented in Ref.~\cite{Jinno:2017fby}. Unfortunately, the method used there was not able to predict the hard part of the spectrum due to oscillations in the multi-dimensional integrals.

Before we contrast our results with hydrodynamic simulations, we qualitatively compare the results from the bulk flow model and the envelope approximation. 
The main difference between the envelope approximation and the bulk flow model is the asymptotic behavior away from the peak of the spectrum. In the ultra-violet ($\omega \gg \beta$), the bulk flow model leads to a scaling $\sim k^{-3}$, while the envelope shows only a slow decent $\sim k^{-1}$. This is due to the fact that the envelope approximation leads to many kinks and cusps in the energy distribution and also in the time profile of the anisotropic stress. The bulk flow model, leads to a rather smooth anisotropic stress that in turn leads to a suppression of GW production for large frequencies~\footnote{Notice that the final scaling $\sim k^{-3}$ is due to a cancellation between the contributions during the phase transition and the contribution from the end of the phase transition up to very late times, that we integrate analytically, see (\ref{eq:infiPiece}). This also serves as an important check.}.
In the infrared ($\omega \ll \beta$), the bulk flow model leads to a scaling $\sim k^1$, while the envelope shows a steeper slope, $\sim k^3$. Naively, the power spectrum has to scale at least as $\sim k^3$ for causally generated spectra~\cite{Caprini:2009fx}, so at first sight the scaling $\sim k^1$ is surprising. Ultimately, The reason is that the source is lasting forever in the bulk flow model. At the same time, the nucleated bubbles keep expanding thus shifting the produced GW signal to smaller frequencies. In effect, this is the reason for the factor $1/\omega$ in (\ref{eq:infiPiece}) that leads to the enhancement of the infrared part of the spectrum. For frequencies $k \ll H$, the GW spectrum will scale as $k^3$ which is however not recovered explicitly since we neglect the expansion of the Universe. Finally, the bulk flow model leads to 
slightly smaller peak frequencies compared to the envelope approximation, which is due to the longer lasting source.

Some behavior we find is in agreement with hydrodynamic simulations. For example, we observe the scaling $k^{-3}$ for large wave numbers that is also seen in simulations.
Nevertheless, we also find qualitative differences. The main finding of the hydrodynamic simulations is that after the phase transition, a constant anisotropic stress persists in the plasma. This is also true in our bulk flow model. However, in the bulk flow model, the stress slowly moves to smaller wave numbers as just explained. In the hydrodynamic simulations the typical correlation length seems to stay constant after the phase transition. This leads to an additional enhancement of the GW power spectrum of order $\beta/H$ (assuming that the Hubble expansion will ultimately damp the anisotropic stress). On the other hand, the produced GW spectrum in our model is finite even without this assumption. Besides, the most recent simulations~\cite{Hindmarsh:2015qta, Hindmarsh:2017gnf} show that the peak of the GW spectrum in hydrodynamic simulations is due to the thickness of the sound shells and not the average bubble size. In the bulk flow model, the spectrum decreases beyond the peak that is related to the average bubble size.

How can the hydrodynamic simulations lead to a qualitatively different result than the bulk flow model? Most approximations in the bulk flow model seem easily justified, at least in the detonation regime~\cite{Jinno:2017fby}, and should not have a major impact on the GW spectrum. One realistic possibility is that after the first collision, the sound shells lose energy by reducing the propagation speed. The bulk flow model explicitly assumes that the propagation speed of the sound shell is unchanged and only the energy density is reduced. In particular, the spherical shape of the nucleating bubbles is preserved while the energy density varies on the surface. Ultimately, a change in propagation speed could lead to a fragmentation of the bubbles and a constant correlation length at late times. 
Another possibility is that collisions of sound shells that happen after the phase transition is finished play an important role. In the bulk flow model, the energy density is a linear superposition of the energy densities of the individual bubbles. Besides, the wall thickness is shrunk to zero. This renders collisions after the phase transition irrelevant. In the hydrodynamic simulation, the velocity field is (in the weak regime) a superposition of the velocity fields of the individual bubbles. Since the anisotropic stress is quadratic in the velocity field, the regions where sound shells overlap might be of importance. 
The last point could also explain why the peak of the GW spectrum is related to the sound shell thickness and not the average bubble size. This line of argument is also supported by the sound shell model~\cite{Hindmarsh:2016lnk} that (assuming a Gaussian velocity field) reproduces many qualitative features of the hydrodynamic simulation.

\section*{Acknowledgments}

First, I would like to express my gratitude to the members of the LISA Cosmology Working Group and to the Mainz Institute for Theoretical Physics (MITP) for hosting its most recent meeting. I am very thankful to Mark Hindmarsh, Stephan Huber, Kari Rummukainen and David Weir for explaining their work in detail to me.
Finally, I would like to thank Iason Baldes for discussions and carefully reading the manuscript. 
I acknowledge support by the German Science
Foundation  (DFG)  within  the  Collaborative  Research  Center  (SFB)  676  Particles,  Strings
and the Early Universe.


\end{document}